# Protein conformational dynamics and electronic structure

Fabio Pichierri[a,b*]

[a]*Department of Applied Chemistry, Graduate School of Engineering, Tohoku University,*

*Aoba-yama 6-6-07, Sendai 980-8579, Japan*

[b]*Quantum Proteomics Initiative (QPI)*

[v.1, August 18, 2013]

**Abstract.**

Quantum mechanical calculations are performed on 116 conformers of the protein ubiquitin (Lange et al., Science 2008, 320, 1471–1475). The results indicate that the heat of formation (HOF), dipole moment, energy of the frontier orbitals HOMO and LUMO, and HOMO-LUMO gap fluctuate within their corresponding ranges. This study thus provides a link between the conformational dynamics of a protein and its electronic structure.

[*] Corresponding author. Tel. & Fax: +81-22-795-4132

*E-mail address:* fabio@che.tohoku.ac.jp (F. Pichierri)

Conformational dynamics represents a fundamental aspect of protein science (and molecular biophysics) which is concerned with the biological function of a protein (Frauenfelder et al., 1991; Grant et al., 2010). At present, the conformational properties of proteins are computationally investigated with the aid of (classical) molecular dynamics (MD) simulations (McCammon and Harvey, 1987; Kukol, 2008; Shaw et al., 2010) whereby chemical bonds and non-covalent interactions (H-bonds, Coulomb and van der Waals interactions, etc.) among atoms are described by model potentials (e.g. the harmonic or Morse potentials for covalent bonds, the Coulomb potential for the interaction between charged atoms or groups, and the Lennard-Jones potential for non-covalent interactions) in combination with force-field parameters such as atom charges, force constants, etc. (Kukol, 2008). Given a potential function, the forces acting on the atoms of a protein can be computed so that they will move according to the Newton equation of motion which is solved numerically for a desired number of time steps.

In spite of the extraordinary success of biomolecular simulations, however, one should not forget that proteins, like any other molecule, are made of positively charged nuclei surrounded by (negatively charged) electrons both of which behave according to the laws of quantum mechanics (Schiff, 1968; Dirac, 1982). It is therefore imperative to establish the relationship between electronic structure and conformational dynamics of proteins. Quantum dynamical calculations based on *semiempirical* hamiltonians can nowadays be employed in the study of solvated proteins, as shown in a recent study by Anisimov and coworkers (Anisimov et al., 2009), whereas *ab initio* quantum dynamical simulations are limited mostly to the study of small peptides (see, for example the study of a Beccara et al., 2011). An alternative approach is represented by *static* quantum mechanical calculations performed on protein conformers obtained either from classical MD simulations (whereby each snapshot represents a conformer along the MD trajectory) or using experimentally determined conformers. In this regard, a few years ago the author investigated for the first time the relation between electronic structure and protein dynamics (Pichierri, 2005) by employing an ensemble of 128 conformers (PDB id 1XQQ) of the protein ubiquitin determined by NMR spectroscopy (Lindorff-Larsen et al., 2005).

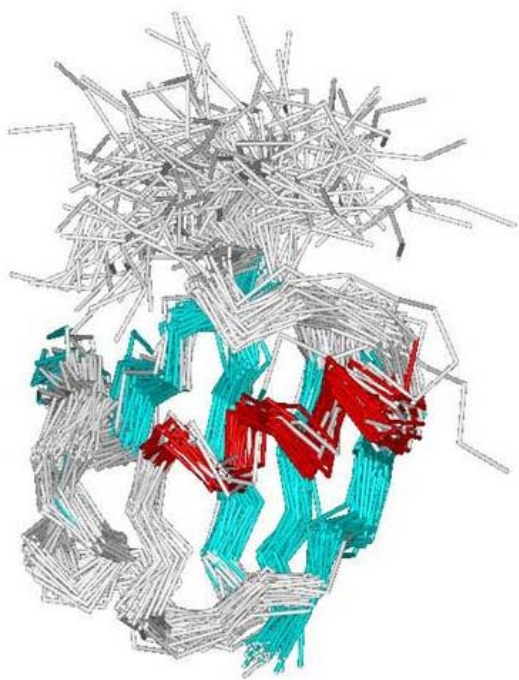

**Figure 1.** Superimposed $C_\alpha$ traces of the 116 conformers of ubiquitin (EROS ensemble, PDB id 2K39).

Here we present a quantum mechanical analysis of an ensemble of 116 ubiquitin conformers whose structures were also determined by NMR spectroscopy (Lange et al., 2008; Ban et al, 2011). The conformers of this ensemble, termed EROS ensemble (PDB id 2K39), are shown in Figure 1. Each ubiquitin conformer (empirical formula: $C_{378}H_{629}N_{105}O_{118}S$ corresponding to 1231 atoms; charge=0) was subjected to a quantum mechanical calculation using the semiempirical PM6 hamiltonian (Stewart, 1996; Stewart, 2007; Stewart, 2008; Stewart, 2009) in combination with the Conductor-like Screening Model or COSMO (Klamt and Schüürmann, 1993) for the implicit treatment of the solvent (the recommended value of 78.4 was employed here for the dielectric constant ε of water). The computed properties (heat of formation, HOF, magnitude of the dipole moment, energy of the highest-occupied and lowest-unoccupied MO levels, HOMO and LUMO, and the HOMO-LUMO energy gap) of this ensemble are shown in Figure 2. Notice that in semiempirical calculations it is customary to quantify the total energy of the system with the corresponding HOF (in kcal/mol) rather than the electronic energy (in hartrees or atomic units) as it is done in the case of *ab initio* calculations on small molecules (Cramer, 2004; Jensen, 2006).

As shown in Figure 2a, the HOF of the conformers' ensemble varies within a range of 135.3 kcal/mol, from −5370.5 kcal/mol (conformer 99) to −5235.2 kcal/mol (conformer 71). This result is consistent with the presence of many rotatable bonds in this protein, each of them contributing a small energy amount to the total change in HOF. The magnitude of the dipole moment (expressed in Debye or D), displayed in Figure 2b, varies from 127 D to 347 D and the average value of the dipole corresponds to ~230 D. Interestingly, the dipole moment vector of the lowest-energy conformer (conformer 99) has a magnitude of 238 D and is oriented from the α-helix to the β-sheet, as shown in Figure 3 and in agreement with a previous study of the author on a different ensemble of conformers (Pichierri, 2005).

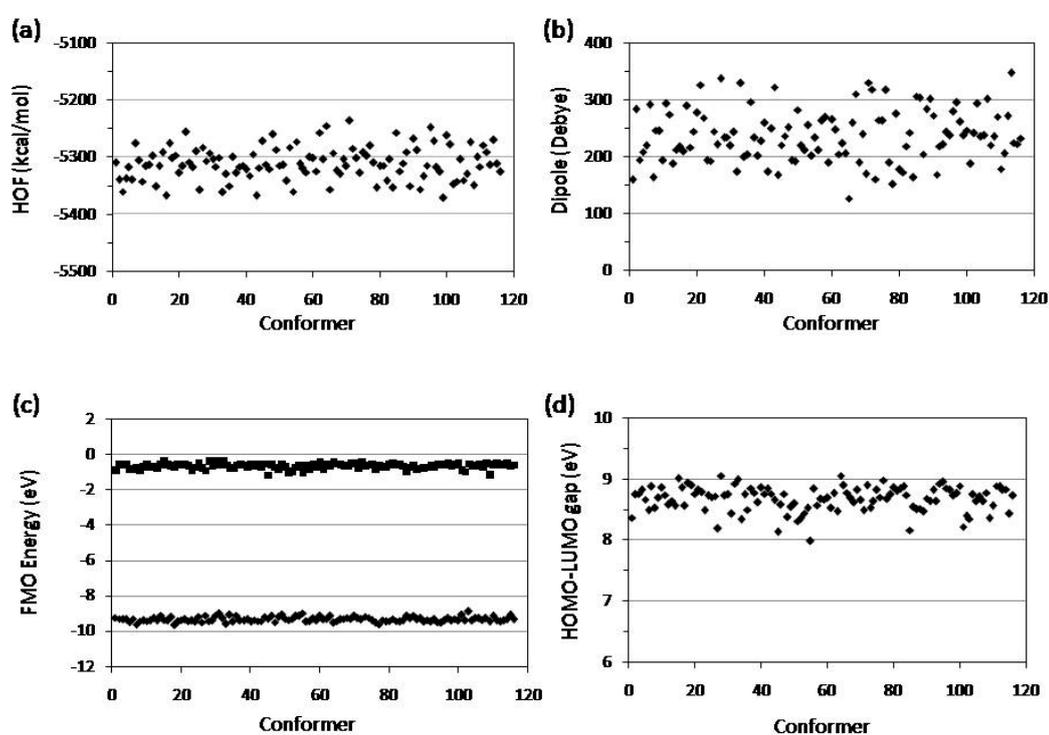

**Figure 2.** Distribution of the computed properties within the EROS ensemble: (a) heat of formation (HOF); (b) dipole moment; (c) HOMO (♦) and LUMO (■) energies; (d) HOMO-LUMO energy gap.

As far as the energies of the frontier MOs are concerned, the HOMO energy ranges from −9.647 eV (conformer 18) to −8.986 eV (conformer 55) while the LUMO energy ranges from −1.173 eV (conformer 45) to −0.331 eV (conformer 32), as shown in Figure 2c. Most interesting, the HOMO-LUMO gap of the conformers' ensemble shown in Figure 2d ranges from 7.992 eV (conformer 55) to 9.057 eV (conformer 28) with the most stable conformer (conformer 99) being characterized by a gap of 8.766 eV. This value is almost three times as large as that of 3.4 eV computed by Payne and coworkers (Lever et al., 2013) for ubiquitin (PDB id 1UBQ) using Kohn-Sham density functional theory (DFT) and an implicit solvent model. With this approach the authors investigated five more proteins whose computed gaps ranged from 3.3 eV to 3.7 eV. The magnitude of the gap obtained from DFT seems somehow underestimated with respect to Hartree-Fock theory, as is also the case for small molecules. In a recent theoretical study on seventeen proteins, Rudberg showed that Hartree-Fock calculations produced gaps in the range from 3.64 eV to 12.03 eV (Rudberg, 2012) whereas the corresponding DFT-computed gaps were considerably smaller in magnitude being in the ranges 0.11–4.16 eV (B3LYP), 0.13–4.65 eV (PBE0), and 0.25–7.23 eV (BHandHLYP). These values of the HOMO-LUMO gaps were obtained in the gas-phase and using the 6-31G** basis set.

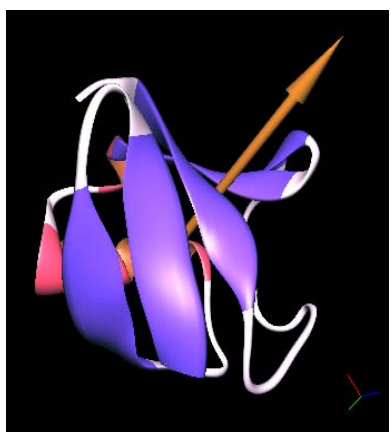

**Figure 3.** Dipole moment vector of the ubiquitin conformer 99 (μ=238 Debye).

Given the impossibility of a comparison with experimental data, however, it is difficult to assess which model chemistry yields the most realistic energy gap. Nevertheless, the results shown in Figure 2 indicate that the HOMO-LUMO gap along with the other electronic properties display some variation as a result of the conformational dynamics of the protein. Importantly, however, while the energy of the MO levels oscillates within the above ranges, the *localization* of the MOs on specific residues of the protein is not affected by the dynamics (Pichierri, 2005). This is in line with the fact that any specific biological function, such as enzymatic reactivity, which is determined by the frontier orbitals of the protein should be maintained regardless of the conformational dynamics as long as the protein operates at constant temperature and is not denaturated by either chemical or physical agents. This is not to say that conformational dynamics does not affect catalysis (Bhabha et al., 2011).

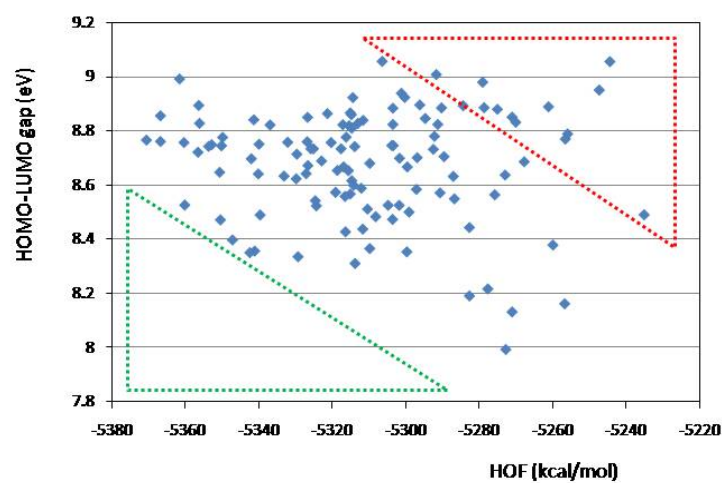

**Figure 4.** Correlation between the HOMO-LUMO gap and heat of formation (HOF) of the 116 conformers of ubiquitin (EROS ensemble).

In Figure 4 is shown the correlation between the HOF (in kcal/mol) and the HOMO-LUMO gap (in eV) of each conformer in the EROS ensemble. Although the data are significantly spread over the graph, therefore making a polynomial fitting unworthy, we notice that the lower left (green) triangular zone does not contain any conformer characterized by both HOF and HOMO-LUMO gap of a large magnitude. On the other hand, the upper right (red) triangular zone with an area equal to the red zone's area contains several conformers which are characterized by relatively low HOF values and high HOMO-LUMO gaps. The graph thus suggests that these two quantities are inversely correlated with each other. At this stage we are not able to fully explain the correlation obtained for the conformers with a relatively low HOF located in the red zone. Also, the non perfect semiempirical potentials employed here along with environmental effects (counterions, explicit solvent, etc.) could play a role in determining the magnitude of the HOF and HOMO-LUMO gap computed for each conformer in the ensemble.

### Acknowledgments

I thank Dr. J.J.P. Stewart (Stewart Computational Chemistry, Colorado Springs) for the continuous developments of the MOPAC software package. This work is supported by the Department of Applied Chemistry of the Graduate School of Engineering, Tohoku University.